# High-capacity reversible hydrogen storage in scandium decorated holey graphyne: Theoretical perspectives

*Vikram Mahamiya[a*], Alok Shukla[a*], Nandini Garg[b,c], Brahmananda Chakraborty[b,c*]*

[a]Indian Institute of Technology Bombay, Mumbai 400076, India

[b]High pressure and Synchrotron Radiation Physics Division, Bhabha Atomic Research Centre, Bombay, Mumbai, India-40085

[c]Homi Bhabha National Institute, Mumbai, India-400094

email: vikram.physics@iitb.ac.in ; shukla@phy.iitb.ac.in ; brahma@barc.gov.in

## ABSTRACT

We have investigated the hydrogen storage capabilities of scandium decorated holey graphyne, a recently synthesized carbon allotrope, by applying density functional theory and molecular dynamics simulations. We have observed that one unit cell of holey graphyne can adsorb 6 Sc atoms, and each Sc atom can adsorb up to 5 $H_2$ molecules with an average binding energy and average desorption temperature of -0.36 eV/$H_2$ and 464 K, respectively. The gravimetric weight percentage of hydrogen is 9.80 %, which is considerably higher than the Department of Energy, United-States requirements of 6.5 %. We have found that a total amount of 1.9e charge transfers from the 3d and 4s orbitals of Sc atom to the C-2p orbitals of holey graphyne by performing the Bader charge analysis. Hydrogen molecules are bonded with the scandium atom by Kubas interactions. The *ab-initio* molecular dynamics simulations confirm the structural integrity of scandium decorated holey graphyne system at the high desorption temperatures. The presence of sufficient diffusion energy barriers for the Sc atom ensure the avoidance of metal-metal clustering in the system.

**GRAPHICAL ABSTRACT**

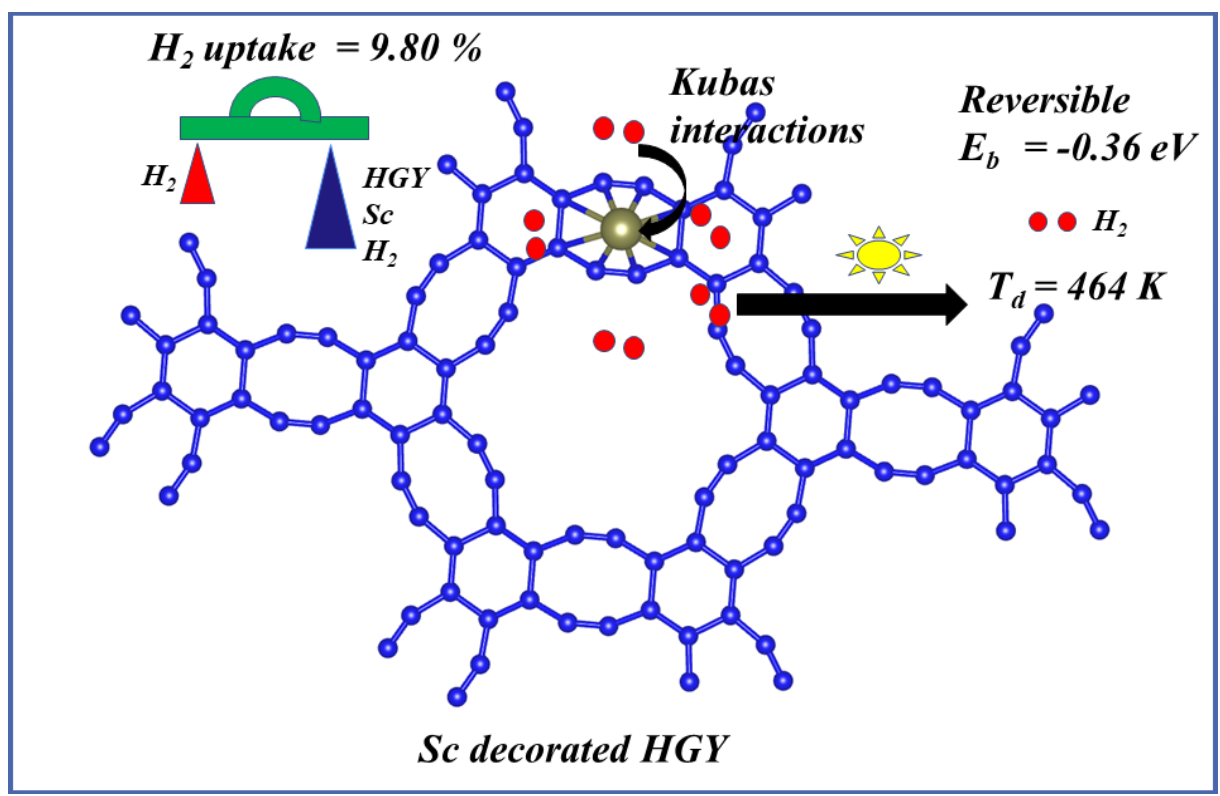

1. **INTRODUCTION**

The global energy demand is increasing day by day, and the conventional fossil fuel energy sources are limited. In addition to that, the combustion of fossil fuel sources produces $CO_2$ and other pollutants. Therefore, it is crucial to find alternative energy sources which can replace fossil fuels in the near future. The scientific community is very optimistic towards hydrogen as a prominent clean energy source as it is highly abundant, recyclable, environmentally friendly, and possesses high energy density compared to any other fuel[1–3]. Safe, compact, affordable, and efficient hydrogen storage are some of the challenges which need to be addressed[4]. Bulky compressed gas tanks are required to store hydrogen in the gas phase at high pressure, but transportation of these gas tanks is costly and involves safety issues. Hydrogen storage is inefficient in the liquid phase due to the high energy cost in liquefaction[1]. A solid-state form of hydrogen storage is suitable if the substrate material can store a sufficient amount of hydrogen molecules for reversible use. Department of Energy, United-States (DoE-US)[5,6] has issued some guidelines[5,6] for the suitable hydrogen storage materials in which the following conditions should be satisfied: (a) The substrate material should adsorb at least 6.5 weight percentage (wt %) of hydrogen[7]. (b) The binding energy of the adsorbed hydrogen molecules should lie in the range of -0.2 eV to -0.7 eV for the practical use of adsorbed hydrogen molecules[8].

Stable Metal hydrides[9–11] can be formed by adsorbing the hydrogen molecules on the metal atom. Hydrogen molecules are bounded with strong chemical bonds; hence poor reversibility of the hydrogen is one of the major concerns in metal hydrides[12]. Different metal alloys, metal organic frameworks, and porous zeolite structures[13–18] are also studied for hydrogen storage. However, these methods are not practically efficient due to the poor reversibility and low uptake of hydrogen.

A lot of research has been carried out on the hydrogen storage properties of carbon nanomaterials. Carbon nanostructures have a large surface area, and the molecular mass of carbon is also relatively small compared to most of the metal, metal alloys substrate. Therefore, different carbon nanostructures are also studied for hydrogen storage applications [19–23]. Hydrogen storage in pristine carbon nanostructures is also practically inefficient at ambient conditions due to the weak van der Waals forces between carbon nanostructures and hydrogen.

Hydrogen molecules are bonded by only the physisorption process; hence the structure is stable only at low cryogenic temperatures[24]. Metal doped carbon nanostructures are proven to be good hydrogen storage media near room temperature[25] because hydrogen molecules are strongly bonded with the metal atom. Hydrogen storage properties of scandium decorated $C_{60}$ and $C_{48}B_{12}$ buckyballs were studied by Zhao et al.[26] and they have found that up to 9 % hydrogen uptake for their systems. Hydrogen storage properties of $Li_6C_{60}$ and $Na_6C_{60}$ were studied by Ren et al.[27]. They have checked the stability of the different isomers of $Li_6C_{60}$ and $Na_6C_{60}$ systems by varying temperature and pressure and using thermodynamic methods. Soltani et al.[28] have explored hydrogen storage in Pd and Co decorated $C_{24}$ fullerene. Hydrogen storage properties of alkali and alkaline-earth metals (Li, Na, K, Mg, Ca) decorated $C_{24}$ fullerene are studied by Zhang and Cheng[29]. They have proposed that a maximum of 12.7 % of hydrogen uptake can be achieved when $C_{24}$ fullerene is decorated with Li atoms. Sathe et al.[30] have studied hydrogen storage in Ti-doped lowest symmetric isomer of $C_{24}$ fullerene. They have found that each Ti atom can bind up to 4 hydrogen molecules leading to 10.5 % of hydrogen uptake. Recently, Mahamiya et al. have reported 13.02 % of hydrogen uptake for Sc decorated $C_{24}$ fullerene[31]. Hydrogen storage in Yttrium doped $C_{24}$ fullerene was also explored by Mahamiya et al.[32]. They have reported 8.84 % of reversible hydrogen uptake, predicting that each Y atom can bind up to 6 $H_2$ molecules. Sahoo et al. have reported 13.08, and 10.80 wt % of hydrogen for Li and Na functionalized $C_{20}$ fullerene[33]. Hydrogen storage in Y doped $B_{40}$ fullerene was studied by Zhang et al.[34]. Hydrogen storage in various kinds of single-walled carbon nanotubes (SWCNTs) (zigzag and armchair) was studied by Tada et. al.[35]. Then hydrogen storage in Ti decorated SWCNT was checked by Yildirim et al.[36]. They have found that Ti can bound 4 molecular hydrogen, which leads to 8 % of hydrogen uptake. The role of different heteroatoms substitution in hydrogen adsorption properties of carbon nanotubes was studied by Sankaran et al.[37]. Hydrogen storage properties of light transition metal (Sc, V) decorated carbon nanotubes and graphene was studied by Durgun et al.[38]. They have found that lighter transition metal bonded to carbon nanotubes can adsorb 5 molecular hydrogens. Hydrogen storage properties of transition metals (Y, Zr, Nb, Mo) decorated SWCNT were studied by Modak et al.[39]. They predicted that transition metal decorated metallic system is better hydrogen adsorber. Chakraborty et al.[40] have studied hydrogen storage in Y decorated SWCNT. They have found that each Y atom can adsorb 6 hydrogen molecules, and the system is stable at extremely high temperatures. Hydrogen storage properties of Ti-doped graphene in the presence of oxygen were studied by Rojas et al.[41]. They have found that when their substrate is exposed to hydrogen only, each metal atom can adsorb 4 hydrogen molecules; however, in the presence of oxygen Ti atom gets oxidized and form titanium oxide. Hydrogen storage properties of calcium doped graphene using plane-wave calculations were studied by Ataca et al.[42]. They have found that up to 8.4 % of hydrogen uptake is possible when Ca atoms are adsorbed to both sides of graphene. Ti-doped graphene for hydrogen storage was studied by Liu et al.[43]. Hydrogen storage in Y decorated pristine and boron-doped graphene was studied by Liu et al.[44]. They have found that Y decorated boron-doped graphene's hydrogen uptake can reach up to 5.78 %. Ti decorated zigzag graphene nanoribbons system was studied for hydrogen storage by Lebon et al.[45]. They have found that each Ti atom can bound 4 molecular hydrogens leading to 6 % of hydrogen uptake for the system. Hydrogen storage in Zr decorated graphene was studied by Yadav et al.[46]. They have found that the desorption temperature of the adsorbed hydrogen molecules increases with the increase in the magnetic moment of the system. Hydrogen storage in Co-doped graphene was investigated by Bakhshi et al.[47] using DFT and DFT-D3[48,49] methods. Hydrogen storage in different metal decorated (Li, Ca, Sc, Ti) graphyne was studied by Guo et al.[50]. Gangan et al.[51] have found that the acetylene linkage in graphyne makes the system nonmagnetic after the decoration of Y atom resulting in higher wt % of hydrogen as compared to the magnetic systems. Hydrogen storage in different transition metals (Sc, Ti, V, Cr, Mn) decorated covalent triazine-based frameworks (CTF) was studied by He et al.[52]. They have reported that Sc and Ti decorated CTF are good hydrogen storage substrates near ambient conditions. Vaidyanathan et al.[53] have reported 7.1 wt % of hydrogen uptake for Zr doped triazine frameworks. Recently, hydrogen storage in Ti-doped psi graphene was studied by Chakraborty et al.[54]. They have found that each Ti can adsorb 9 hydrogen molecules leading to 13.1 wt % hydrogen for their system.

Hydrogen storage in Pd nanoparticle-doped multi-walled carbon nanotubes (MWCNTs) was studied by Mehrabi et al.[55] experimentally. They have used the laser ablation method and achieved 6 % of hydrogen uptake. Tarasov et al.[56] have studied hydrogen storage in the nanocomposites of Mg with Ni and graphene-like material experimentally and found that reversible hydrogen storage higher than the DoE requirement can be achieved.

Graphyne and its analogous are proven to be good energy storage substrates. Srinivasu et al.[57] have studied the Li dispersed graphyne and graphdiyne for lithium and hydrogen storage. Li and Na decorated graphdiyne are studied for hydrogen storage by Wang et al.[58]. They have found that hydrogen storage uptake can reach up to 8.81 % and 7.73 % for Li and Na decorated graphdiyne, respectively. Panigrahi et al.[59] have studied hydrogen storage properties of nitrogenated holey graphene by selective decoration of $Ti_n$ (n = 1 to 5) clusters. Hydrogen storage capabilities of $Li^+$, $Na^+$, $Mg^{2+}$, and $Ca^{2+}$ decorated nitrogenated holey graphene and biphenylene carbon structures were investigated by Guerrero-Aviles et al.[60]. They have found that reversible hydrogen storage is possible in cation decorated holey graphene and biphenylene carbon structures.

Recently, a 2d graphyne analogous material, holey graphyne (HGY), was successfully synthesized by Liu et al.[61]. In HGY, benzene rings are connected by acetylene linkage forming uniform holes which makes them suitable for energy storage applications. In HGY six and eight vertex carbon rings are connected and the ratio of sp to $sp^2$ carbon is 0.5. HGY is a p-type semiconducting material with high hall and electron mobility. Recently, the hydrogen storage properties of Li decorated HGY were investigated by Gao et al.[62]. They have found that the gravimetric density of hydrogen storage for Li decorated HGY can reach up to 12.8 %, but the issue is that the desorption temperature is less than the room temperature. Hydrogen storage properties of transition metal decorated HGY have not been studied yet to the best of our knowledge. Sc is the lightest transition metal which has a sufficient number of empty d-orbitals to bind the hydrogen molecules with Kubas interaction[63]. In addition to that, the Sc decorated carbon nanostructures are proven to be good hydrogen storage material[26,38].

Here, we present the hydrogen adsorption and desorption properties of the Sc decorated HGY structure by employing DFT simulation. According to our calculations, Sc decorated HGY can adsorb up to 5 hydrogen molecules in the molecular form, which leads to a high weight percentage (9.80 %) of hydrogen. We have found that the average binding energy of the hydrogen is -0.36 eV, and the average desorption temperature is 464 K, which is very suitable for the reversible use of hydrogen for fuel cell applications. We have discussed the binding mechanism, orbital interactions, electronic structure, and charge transfer process in Sc decorated on HGY and $H_2$ adsorbed on Sc decorated HGY systems, using the partial density of states (PDOS), Bader charge analysis[64] and spatial charge density difference plots. Diffusion energy barrier and molecular dynamics calculations were performed to ensure the absence of metal-metal clustering and integrity of the structure at high desorption temperature. This is one of the important aspects of our work because in most of the previous reports on hydrogen storage diffusion energy barrier and molecular dynamics calculations are not provided. Since Sc decorated HGY has suitable binding energy, desorption temperature, hydrogen uptake, and the system is stable at desorption temperature, we believe that Sc decorated HGY is a practically suitable, high-capacity reversible hydrogen storage material.

## 2. METHODOLOGICAL DETAILS

We have performed the density functional theory (DFT) and *ab-initio* molecular dynamics (AIMD) calculations using Vienna Ab Initio Simulation (VASP) package[65–68]. Generalized gradient approximation (GGA) with Perdew-Burke-Ernzerhof (PBE)[69] exchange-correlation functional was used for the calculations. We have used one unit cell of HGY containing 24 carbon atoms for the calculations, and have taken 15 Å of vacuum space to avoid the periodic interactions between the different layers of HGY in our calculations. The Brillouin zone sampling was done using a Monkhorst-Pack *k*-grid of the size 5*5*1. For the plane-wave basis expansion, we have taken the kinetic-energy cut-off to be 500 eV. For the Hellman-Feynman forces and energy, the convergence limit is set to be 0.01 eV/ Å and $10^{-5}$ eV, respectively. As we know that there are weak van der Waals (vdW) interactions present in the system, we have corrected our DFT-GGA results by employing Grimme's dispersion corrections of the DFT-D2[48,49] type. We have also performed the hybrid density functional theory calculation using sophisticated HSE06 functional[70] to compare the binding energy of adsorbed hydrogen molecules with the GGA+DFT-D2 method. We have also performed the *ab-initio* molecular dynamics simulations (AIMD)[71] for HGY + Sc and HGY + 2 Sc composites to check the structural solidity at the desorption temperature. The molecular dynamics simulations were performed by keeping the system in the canonical ensemble for 5 ps time duration in the time step of 1 fs at a constant temperature of 500 K.

## 3. RESULTS AND DISCUSSIONS

### 3.1 Interaction of Sc on HGY

The geometrically optimized structure of 2*2*1 supercell of HGY is presented in **Fig. 1(a).** HGY has a hexagonal lattice structure with the space group P6/mmm. The lattice constants for the HGY unit cell were found to be a = b =10.85 Å, excellently matching with the literature[62]. There are two types of carbon atoms in HGY named C1 and C2 with $sp^2$ and sp hybridization. The unit cell of HGY containing 24 carbons is presented in **Fig. 1(b).** There are 4 different types of C-C bond lengths in the HGY structure, as shown in **Fig. 1(b).** The bond B1 is in between sp-sp hybridized carbon atoms with a bond length of 1.23 Å. B2 is in between sp-$sp^2$ hybridized carbon having 1.41 Å bond length, then there are two $sp^2$-$sp^2$ hybridized bonds B3 and B4, with bond lengths 1.46 Å and 1.40 Å, respectively. The bond angle in between acetylene linkage and hexagon is 126.03°, which is a little bit higher than the similar bond angle for graphyne (120°). The structural parameter bond lengths and bond angle are in excellent agreement with the previously reported values[62].

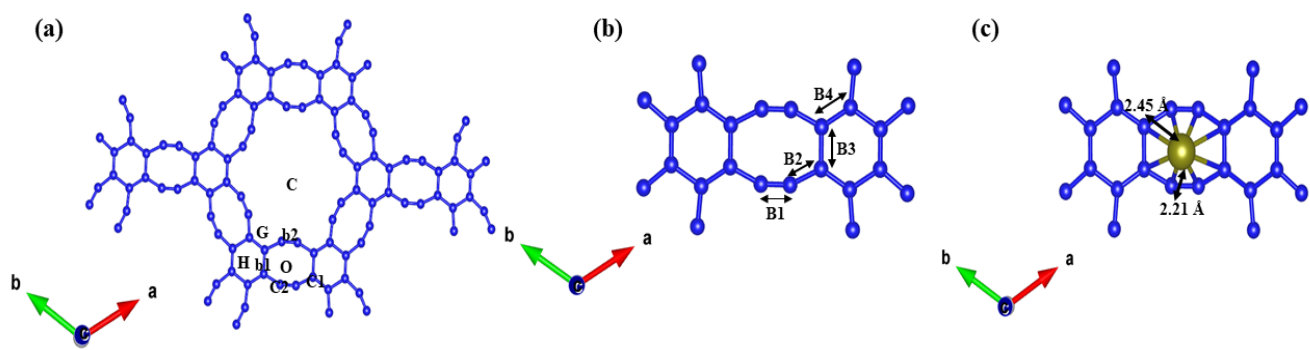

**Fig. 1 Optimized structures of (a) 2*2*1 supercell of HGY (b) Unit cell of HGY with different type of bonds B1, B2, B3 and B4 (c) Sc-decorated HGY where Sc-atom is placed**

**on the top of the octagon of HGY. Blue and golden colors correspond to C-atom and Sc-atoms, respectively.**

After getting the relaxed structure of HGY, we have placed the Sc atom on the top of the different positions of HGY (H, O, C1, C2, b1, b2, G, C) as shown in **Fig.1(a),** at almost 2 Å distance above to the plane of HGY and calculated the binding energy of the attached scandium. We have found that the O position for the Sc atom**,** which is above the center of the octagon of HGY, is the most stable position with maximum binding energy, and scandium placed at b1, C1, C2 also comes at the O position after the relaxation. The optimized structure of Sc decorated HGY at the O position is shown in **Fig. 1(c).** The binding energy of the Sc atom on the top of the O position of HGY is -4.56 eV. The binding energy of the Sc at O position of HGY (4.56 eV) is more than the experimental cohesive energy of Sc (3.90 eV)[72] hence clustering of Sc atom should not take place. In the relaxed structure of Sc decorated HGY, the distance of the Sc atom from the sp and $sp^2$ hybridized carbon atoms are 2.21 Å and 2.45 Å, respectively, while the change in the bond lengths B1, B2, B3, B4 are negligible. Since the O position of the Sc corresponds to the maximum binding energy, we have considered this position for the hydrogen storage. The binding energy of the Sc is calculated using the following equation:

$$\boldsymbol{B.E.(Sc) = E\;(HGY + Sc) - E\;(HGY) - E\;(Sc)} \qquad (1)$$

Where ***B.E (Sc)*** is the binding energy of the scandium placed on the top of HGY. ***E (HGY + Sc), E (HGY),*** and ***E (Sc)*** are the total energies of Sc decorated HGY, pristine HGY and isolated Sc atom, respectively.

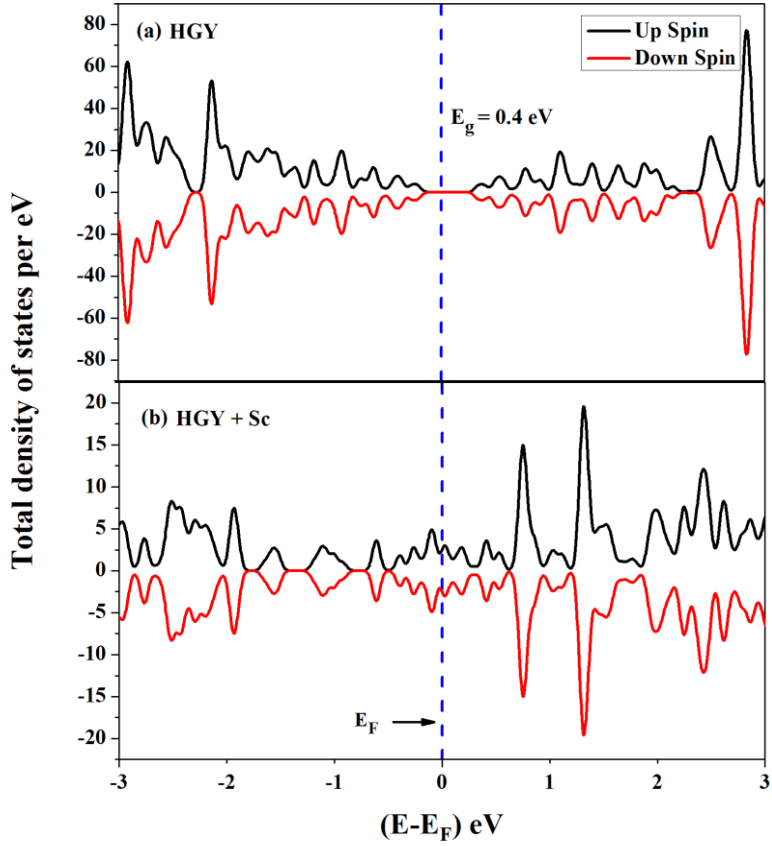

**Fig. 2** Total density of states of (a) HGY (b) Sc-decorated HGY calculated using PBE-GGA level of theory. $E_g$ is the energy band gap. Fermi level ($E_F$) is set at 0 eV.

Next, we are presenting the total density of states of HGY and Sc decorated HGY structures in **Fig. (2).** The energy band gap of HGY is 0.4 eV with PBE-GGA exchange-correlation, which is in good agreement with the reported value of 0.5 eV by Liu et al.[61] calculated using PBE-GGA exchange-correlation. GGA exchange-correlation underestimates the band gap of materials so we have also calculated the band gap of HGY structure using Heyd-Scuseria-Ernzehof (HSE06)[70] exchange-correlation functional. The band gap of HGY with HSE06 exchange-correlation is found to be 0.9 eV, which matches reasonably with the experimental band gap of HGY (~1 eV)[61].

Here, we can notice the presence of the states at Fermi energy in **Fig. 2(b),** which indicates that the HGY structure becomes metallic after the decoration of the scandium atom. Modak et al.[39] have reported that the metallic carbon nanostructures adsorb more hydrogen molecules by Kubas interactions as compared to semiconducting carbon nanostructures. In **Fig. 3 (a & b),** we present the band structure of the HGY and HGY + Sc structures. The band structure of single-layer HGY matches excellently with the literature[61].

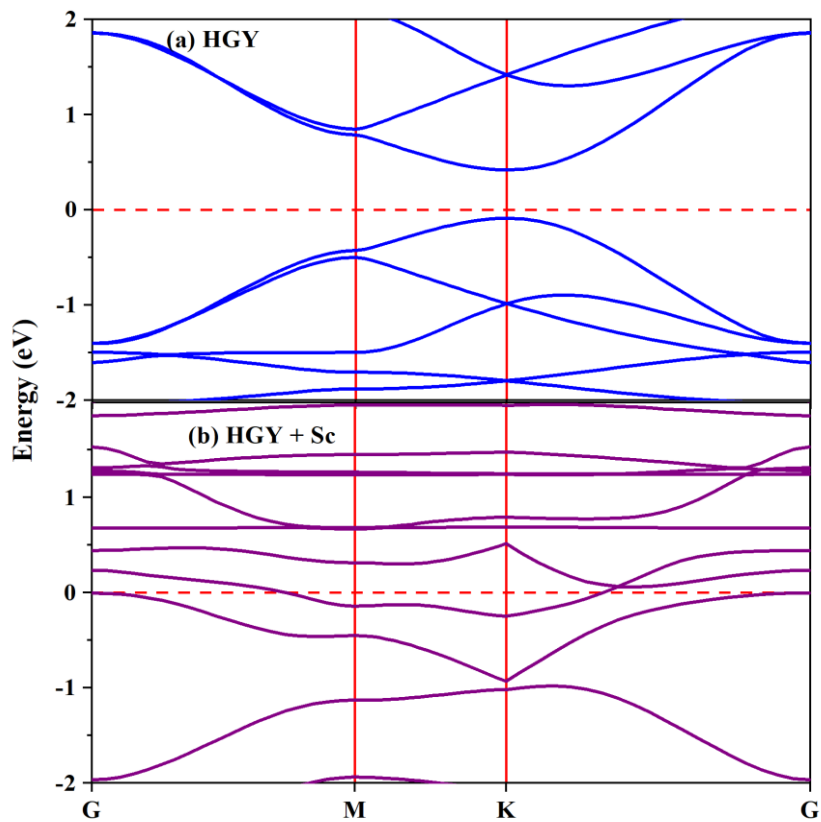

**Fig. 3 Band structure of (a) HGY and (b) Sc decorated HGY calculated using PBE-GGA exchange-correlation. HGY is a semiconducting material having direct energy band gap 0.4 eV. HGY structure becomes metallic after decoration with Sc atom.**

To understand the bonding mechanism and charge transfer phenomenon between the Sc atom and HGY layer, we have plotted the partial density of states, spatial charge density difference plots and performed the Bader charge analysis[64].

**Partial density of states (PDOS) analysis**

To explore the orbital interactions, bonding mechanism, and charge transfer in between the Sc atom and single layer of HGY, we have plotted the partial density of states of C-2p orbitals of pristine HGY and Sc decorated HGY structures as displayed in **Fig. 4(a)** and **Fig. 4(b)** respectively. In **Fig. 4(b),** there are some energy states near the Fermi level, which were absent in the PDOS of C-2p orbitals of pristine HGY displayed in **Fig. 4(a).** The presence of the states near the Fermi level denotes the transfer of charge from the Sc atom to the HGY layer.

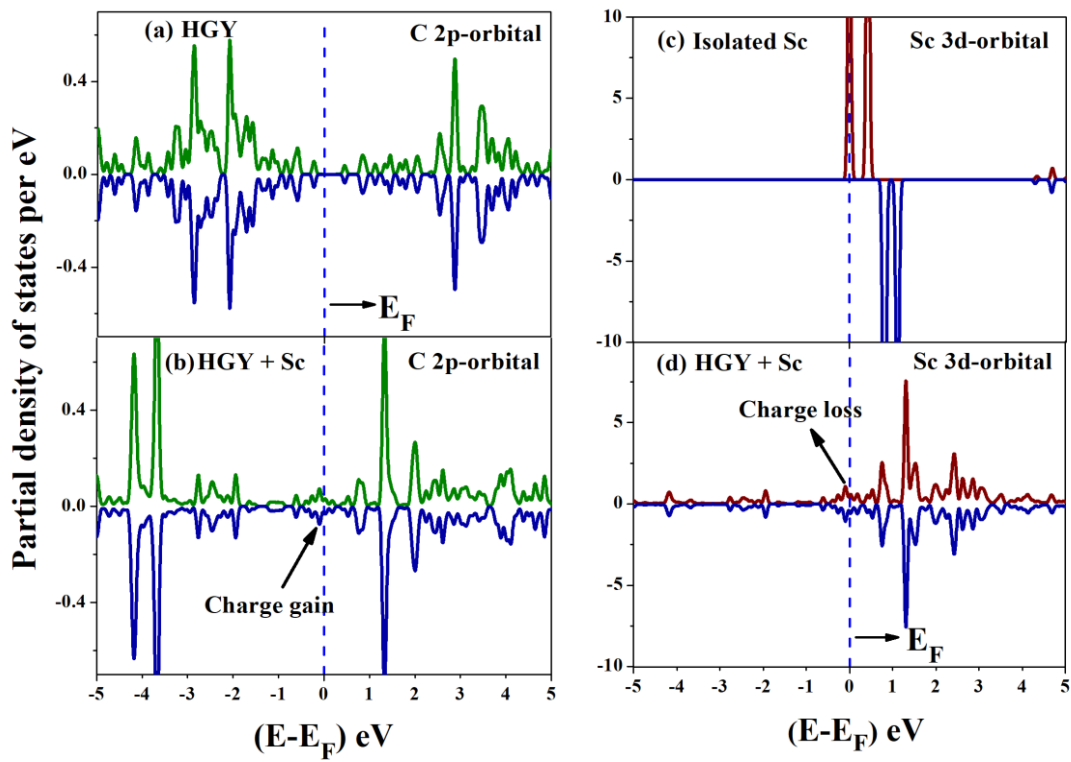

**Fig. 4** Partial density of states for (a) C-2p orbitals of HGY. (b) C-2p orbitals of Sc

**decorated HGY. (c) Sc-3d orbitals of isolated Sc atom. (d) Sc-3d orbitals of Sc decorated HGY. Fermi level is set at zero energy value.**

To get a clearer picture of the orbital interactions and charge transfer, we have plotted the partial density of states of Sc-3d orbitals for isolated Sc, and Sc decorated HGY structure as presented in **Fig. 4(c)** and **Fig. 4(d),** respectively. We can notice that there are some intense states at the Fermi level for isolated Sc atom as shown in **Fig. 4(c),** whose intensity decreases significantly for Sc decorated HGY structure as displayed in **Fig. 4(d).** This is due to the transfer of charge from the Sc-3d orbitals to the C-2p orbitals of HGY when Sc is decorated on the HGY structure. We have also plotted the PDOS of Sc-4s orbital for isolated Sc atom as well as when Sc is decorated on the top of HGY as displayed in **Fig. 5 (a & b).** Here also, the absence of the states near the Fermi level of **Fig. 5(b)** denotes that some charge loss takes place when Sc is attached to HGY. Therefore, from the PDOS plots, it is clear that some charge has been transferred from the valence orbitals of the Sc to the valence orbitals of HGY. This charge transfer is responsible for the binding of the Sc atom (**B.E. =-4.56 eV**) on the HGY structure.

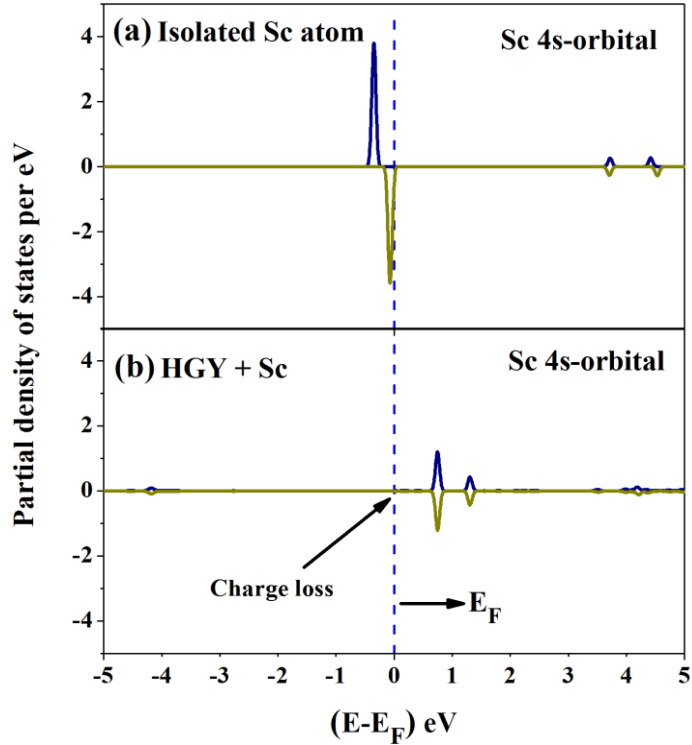

**Fig. 5** Partial density of states for (a) Sc-4s orbital of isolated Sc atom. (b) Sc-4s orbital of Sc decorated HGY. Fermi energy is set at zero energy value.

**Bader charge analysis**

To understand the charge transfer phenomenon quantitatively, we have performed the Bader charge analysis[64]. The Bader charge calculations reveal that a total amount of 1.9e charge has been transferred from the Sc-3d and 4s orbitals to the C-2p orbitals of the HGY unit cell. Due to this significant amount of charge transfer, the Sc atom is bounded strongly on the top of the HGY monolayer.

**Charge density difference plots**

To visualize the charge transfer phenomenon, we have plotted the spatial charge density difference plots as displayed in **Fig. 6.** The top and side view plots of the charge density difference ρ (HGY + Sc) – ρ (HGY) are presented in **Fig. 6(a)** and **Fig. 6(b). Fig. 6(a & b)** are plotted for the iso-surface value 0.0042.

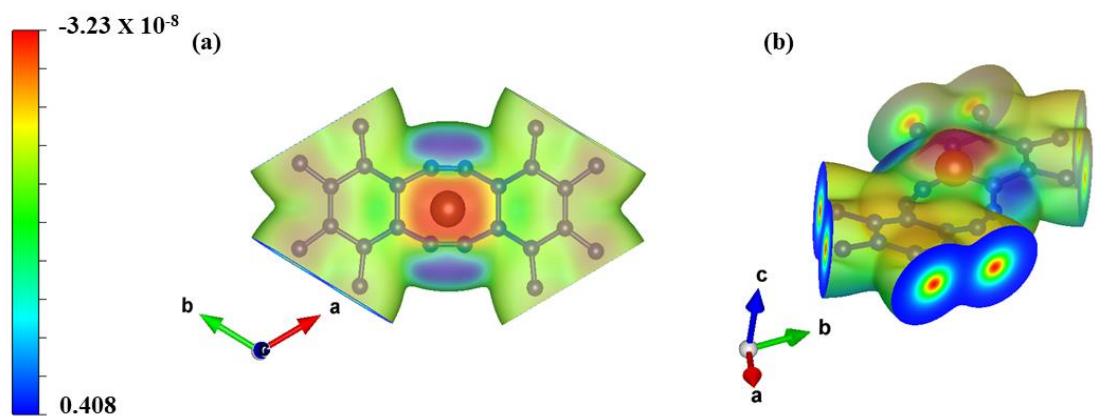

**Fig. 6 Electronic spatial charge density difference plots for (a) Top view of ρ (HGY + Sc) – ρ (HGY) (b) Side view of ρ (HGY + Sc) – ρ (HGY) systems for iso-surface value 0.0042e. The plots are in B-G-R color pattern. Here red color denotes charge loss while green and blue colors denote less and more charge gain regions respectively.**

The charge density difference plots are according to the B-G-R color pattern, in which the red color in the vicinity of the Sc atom denotes the charge loss region. In contrast, the green and blue colors represent less and more charge gain regions, respectively. The charge transfer from the Sc atom to the HGY layer is clear from the charge density plots.

### 3.2 Hydrogen adsorption on Sc decorated HGY

Hydrogen molecules were placed at almost 2 Å distance above the Sc atom of HGY + Sc composition, and then geometry optimization calculations were performed. We have employed the PBE-GGA exchange-correlation for the DFT calculations and Grimme's dispersion method DFT-D2 to incorporate the weak van der Waals interactions. The binding energy of the 1$^{st}$ hydrogen molecule is found to be -0.38 eV, which lies in the suitable range as guided by DoE-US. The binding energy of the 1$^{st}$ hydrogen molecule using sophisticated HSE06 exchange-correlation functional employed in hybrid density functional theory is found to be -0.37 eV, which matches excellently with GGA+DFT-D2 result of -0.38 eV. Therefore, the adsorption energy calculations are accurate and reliable. The H-H bond length of the 1$^{st}$ adsorbed hydrogen molecule changes from 0.74 Å to 0.77 Å. We have found that one Sc atom can adsorb a maximum number of 5 hydrogen molecules with the binding energy in the range -0.2 eV to -0.7 eV as guided by DoE-US. The average binding energy of all 5 hydrogen molecules is found to be -0.36 eV and -0.32 eV, calculated using GGA+DFT-D2 and HSE06 exchange-correlations functional, respectively, which is in between the range specified by DoE, hence very suitable for practical fuel cell applications.

The binding energy of the adsorbed hydrogen molecule was calculated using the following equation:

$$B.E. \text{ of } n^{th} H_2 = E(HGY + Sc + nH_2) - E(HGY + Sc + (n-1)H_2) - E(H_2) \qquad (2)$$

Here n is the total number of hydrogen molecules attached in the current step of calculations. The binding energies of all hydrogen molecules are presented in **Fig. 7.**

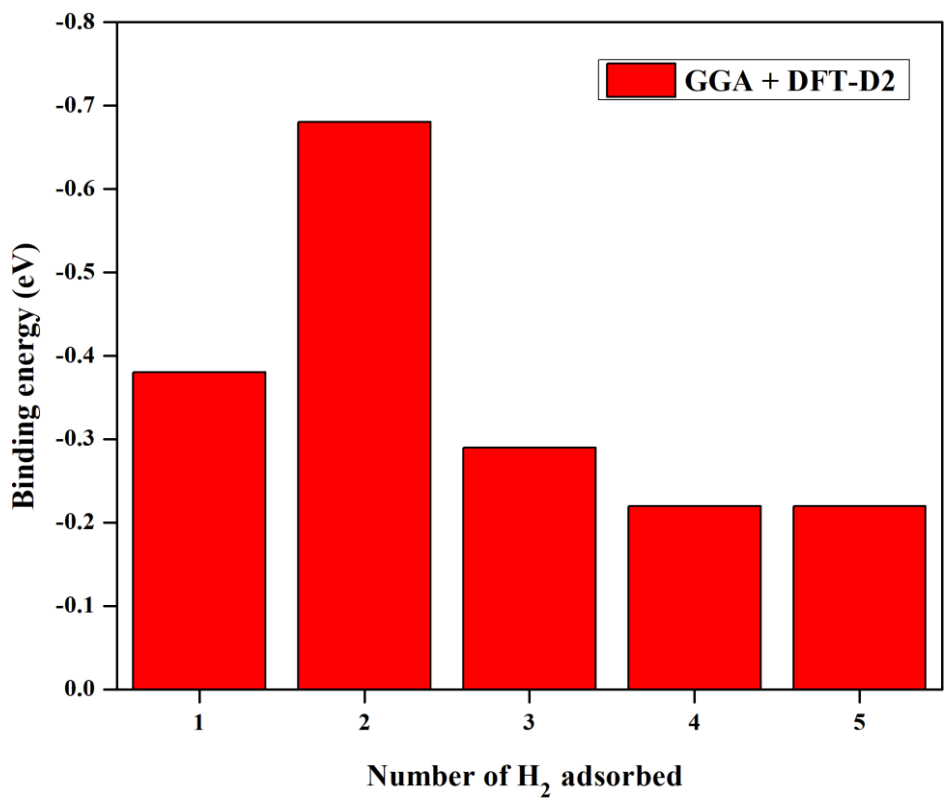

**Fig. 7** Binding energy of the adsorbed hydrogen molecules on Sc decorated HGY using GGA exchange-correlation along with the DFT-D2 dispersion corrections. The binding energies of the hydrogen molecules lie in the suitable range as specified by DoE-US.

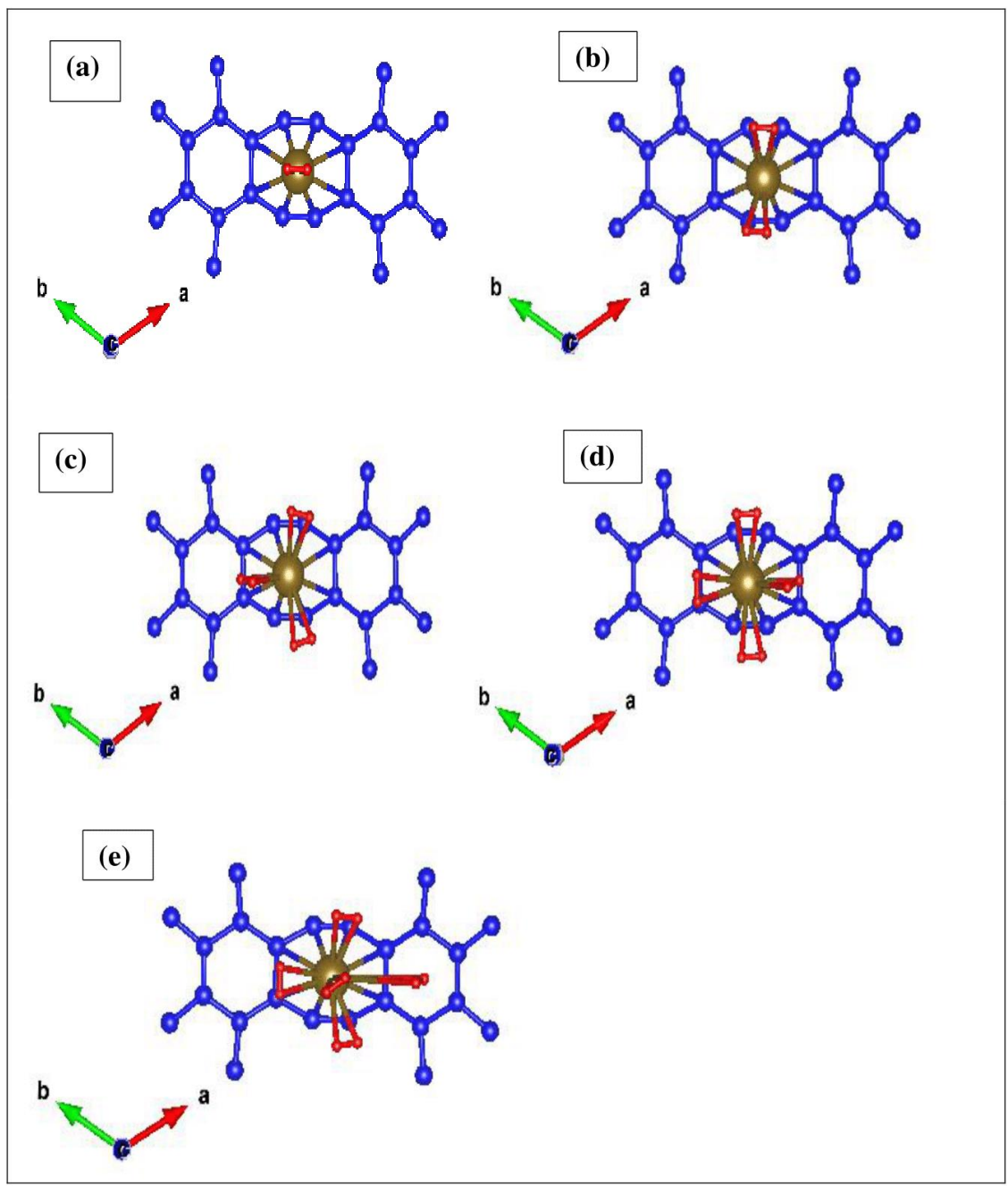

**Fig. 8** Optimized structure of (a) HGY + Sc + H$_2$ (b) HGY + Sc + 2H$_2$ (c) HGY + Sc + 3H$_2$ (d) C$_{24}$ + Sc + 4H$_2$ and (e) HGY + Sc + 5H$_2$ compositions. Blue, golden, and red color spheres denote carbon, scandium and hydrogen atom, respectively.

The optimized structures of HGY + Sc + H$_2$, HGY + Sc + 2H$_2$, HGY + Sc + 3H$_2$, HGY + Sc + 4H$_2$, and HGY + Sc + 5H$_2$ are presented in **Fig. 8.** The binding energies of the attached hydrogen molecules along with the Sc-H and H-H bond lengths after the attachment of hydrogen molecules are presented in **Table 1.**

**Table 1. Average adsorption energy and bond distances (Sc-H and H-H) of the adsorbed H$_2$ molecules with DFT GGA + DFT-D2 and hybrid DFT (HSE06) methods.**

| Compositions | Binding energy (eV) GGA + DFT-D2 | Binding energy (eV) HSE06 | Bond length (Å) Sc-H | Bond length (Å) H-H |
|---|---|---|---|---|
| HGY + Sc + H$_2$ | -0.38 | -0.37 | 2.31 | 0.77 |
| HGY + Sc + 2H$_2$ | -0.68 | -0.56 | 2.33 | 0.76 - 0.77 |
| HGY + Sc + 3H$_2$ | -0.29 | -0.26 | 2.41 | 0.76 - 0.77 |
| HGY + Sc + 4H$_2$ | -0.22 | -0.22 | 2.51 | 0.76 - 0.77 |
| HGY + Sc + 5H$_2$ | -0.22 | -0.20 | 3.93 | 0.75 - 0.77 |
| Average binding Energy per H$_2$ | -0.36 | -0.32 | | |
| Average desorption Temperature | 464 K | 412 K | | |
| Gravimetric wt % | 9.8 % | | | |

We have noticed that one hydrogen molecule can spill to the nearest hexagon site (H) of HGY when five hydrogen molecules are attached to a single Sc atom, as shown in **Fig. 8(e).** This spillover may lead to some interaction between hydrogen molecules attached to two adjacent Sc atoms, affecting the average adsorption energy of hydrogen molecules and, therefore, weight percentage. To confirm that the average binding energy is still in the optimal range, we have performed a geometry optimization calculation on 2*2*1 supercell of HGY, containing two Sc atoms at adjacent O sites of HGY, and 5 hydrogen molecules are attached on each Sc atom. The relaxed structure of this configuration (HGY + 2Sc + 10 H$_2$) is shown in **Fig. 9.**

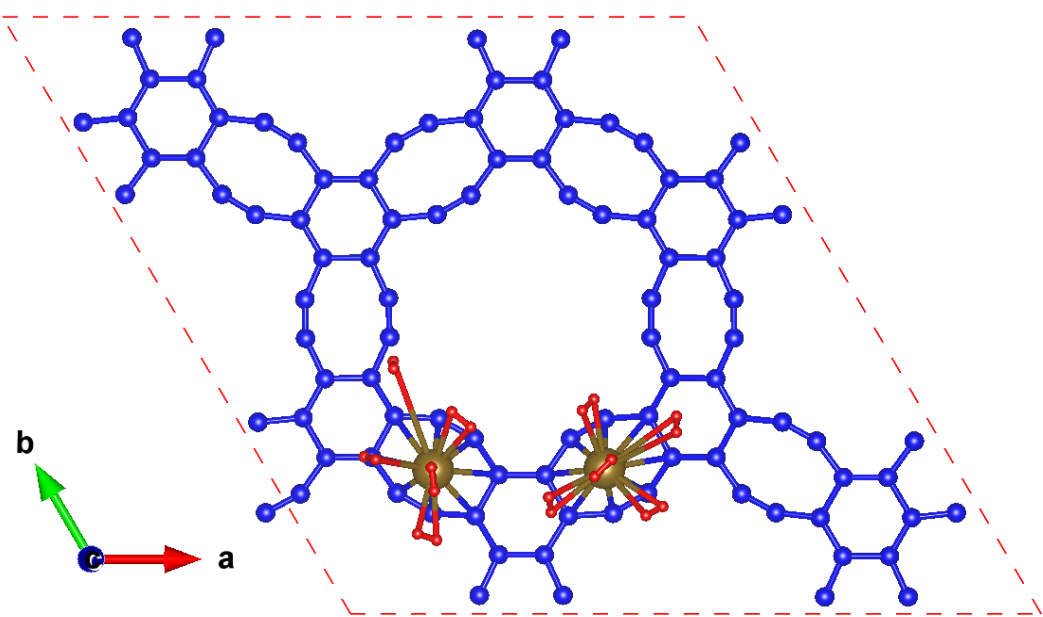

**Fig. 9 Optimized structure of 2*2*1 supercell of HGY containing 2 Sc atoms at the adjacent O sites. 5 hydrogen molecules are attached to each Sc atom. Blue, golden and red color spheres denote carbon, scandium, and hydrogen atom, respectively.**

We have calculated the average binding energy of 10 $H_2$ molecules attached to two adjacent Sc atoms, which is -0.32 eV, in good agreement with the average binding energy of 5 $H_2$ molecules attached to a single Sc atom (-0.36 eV). So, the interaction between the hydrogen molecules attached to two adjacent O sites is negligible, and the average binding energy is in the optimal range for hydrogen storage.

### 3.3 Interaction of hydrogen with the Sc decorated HGY

To explain the orbital interactions and charge transfer between the adsorbed hydrogen molecules and the Sc atom, we have plotted the partial density of states plots and performed the Bader charge analysis. We have also explained the nature of interactions present in the system when hydrogen molecules are attached.

**Kubas-type of interactions**

We have found that the average binding energy of attached hydrogen molecules on the Sc decorated HGY structure is -0.36 eV. The value of the average binding energy lies in between the physisorption and chemisorption processes. We can also notice in **Table 1,** that the H-H bond lengths increases from 0.74 Å to 0.76 Å – 0.77 Å for different hydrogen molecules. This small elongation in H-H bond length is due to the Kubas type of interactions[63,73,74], in which the change in H-H bond length is small, and the hydrogens remain intact in the molecular form. In Kubas interactions, some charge transfers from the filled highest occupied molecular orbitals (HOMO) of the hydrogen molecules to the vacant 3d orbitals of the transition metal (Sc), and subsequently, there is some back charge transfer from the filled 3d orbitals of transition metal to the empty lowest unoccupied molecular orbital (LUMO) of hydrogen molecule take place. In this process, a small amount of charge is gained by the hydrogen molecule, which is responsible for the binding of hydrogen from the scandium atom and the H-H bond length elongation. We have plotted the charge density of the molecular orbitals of the extended HGY, Sc decorated HGY, HGY + Sc + 1$H_2$, and HGY + Sc + 2$H_2$ compositions, presented in **Fig. S1** of the supporting information file. There is a strong interaction between the dumbbell shaped 3d orbitals of Sc atom and $\pi$ bonding orbitals of HGY, as shown in **Fig. S1 (c & d)**. In contrast, the interaction between H 1s orbital and Sc 3d orbitals is small, which is Kubas interaction shown in **Fig. S1 (e, f, g & h)**. Due to Kubas interaction, the spherical shape of the 1s orbitals of the H atoms gets distorted. The molecular orbital charge density plots for the HGY + Sc + 2$H_2$ composition is also similar to the HGY + Sc + 1$H_2$ composition as shown in **Fig. S1 (g & h),** which indicates that hydrogen molecules are not interacting with each other. There are not much changes in the molecular orbital charge density plots for 3$H_2$, 4$H_2$, and 5$H_2$ addition to Sc decorated HGY structure compared to HGY + Sc + 1$H_2$ and HGY + Sc + 2$H_2$ compositions, so we have not included them in the manuscript.

**Partial density of states (PDOS) analysis**

To understand the orbital interactions, charge transfer, and bonding mechanism in between the adsorbed hydrogen molecules and the Sc decorated HGY structure, we have plotted the PDOS of H – 1s orbital for isolated hydrogen molecule and HGY + Sc + $H_2$ compositions as presented in **Fig. 10 (a & b).**

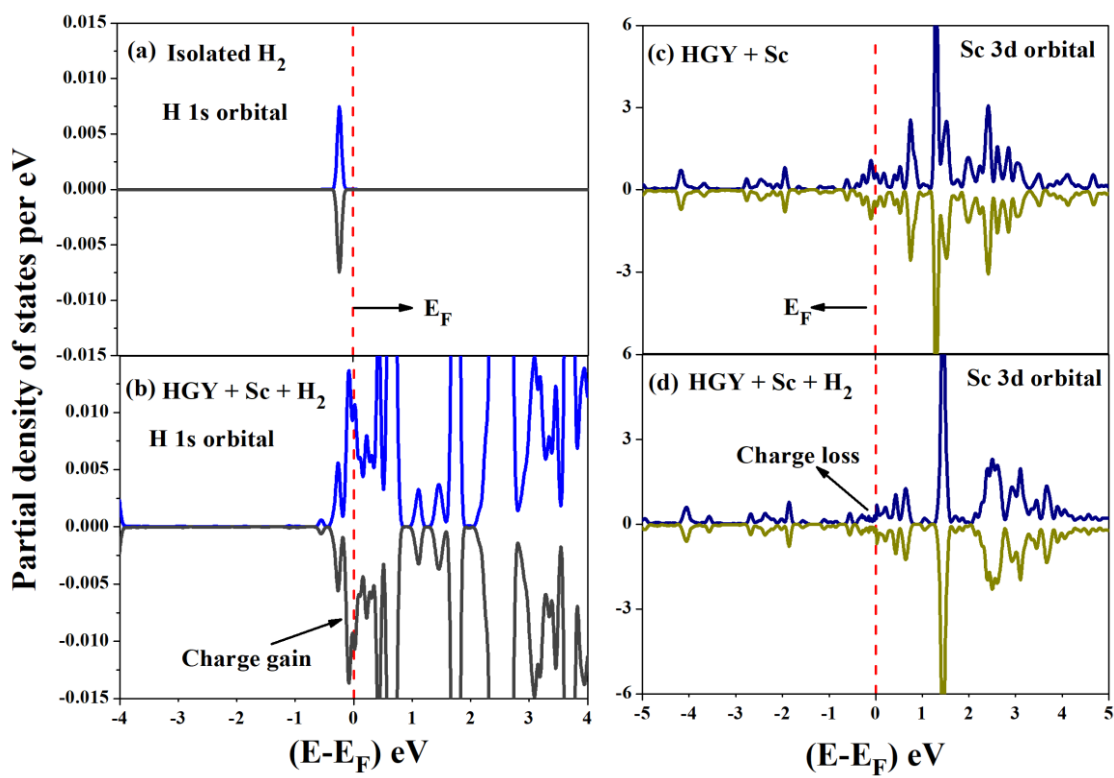

**Fig. 10** Partial density of states for (a) H-1s orbital of isolated $H_2$ molecule. (b) H-1s orbital for HGY + Sc + $H_2$. (c) Sc-3d orbitals of HGY + Sc. (d) Sc-3d orbitals of HGY + Sc + $H_2$. Fermi energy is set at zero energy value.

Here we can see the enhancement in the states near the Fermi level in **Fig. 10 (b),** which indicates that some charge has been transferred from the Sc-3d orbitals to 1s orbital of hydrogen. To confirm this charge transfer process, we have also plotted the PDOS of Sc-3d orbitals for HGY + Sc and HGY + Sc + $H_2$ compositions as displayed in **Fig. 10 (c & d).** From **Fig. 10 (c & d),** we can notice some minor loss in the states of the Sc-3d orbitals, near to the Fermi level for HGY + Sc + $H_2$ system. Hence, we can conclude that there is some small charge transfer take places from the Sc-3d orbitals to the 1s orbital of the attached hydrogen molecule.

This charge transfer is responsible for the binding of hydrogen from the Sc-atom. We have also plotted the total density of states for the addition of hydrogen molecules in the Sc decorated HGY structure. The total density of states of HGY + Sc + nH$_2$ for (n = 1 to 5) are presented in **Fig. S2.** There are not much changes in the total density of states of the Sc decorated HGY system with the addition of H$_2$ molecules. Although the total density of states of the Sc decorated HGY system is slowly decreasing near the Fermi energy with the addition of hydrogen molecules, as shown in **Fig. S2**, the metallic Sc decorated HGY structure remains metallic up to all five hydrogen molecules addition.

**Bader charge analysis**

Bader charge calculation was performed for HGY + Sc and HGY + Sc + H$_2$ compositions to get the amount of charge transferred from Sc to H$_2$ molecule. We have calculated that a total amount of 1.83e and 0.08e charge has been transferred from the Sc atom to the HGY and H$_2$ molecule of HGY + Sc + H$_2$ composition, respectively. The amount of charge transferred from the Sc to H$_2$ molecule is small (0.08e), due to which small elongation in the H-H bond length of the attached hydrogen molecule takes place.

### 3.4 Gravimetric weight percentage (wt %) of the hydrogen

Sc atom's arrangement for the weight percentage calculations is shown in **Fig. 11.** We have placed 6 Sc atoms in one unit cell of HGY, as shown in **Fig. 11 (a & b).**

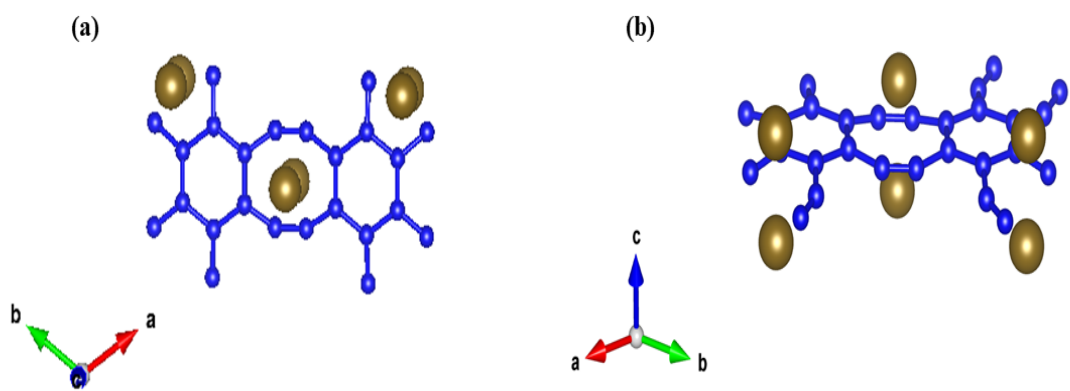

**Fig. 11 Metal loading pattern for HGY unit cell (a) Top view of HGY + 6 Sc composition (b) Side view of HGY + 6 Sc composition. Gravimetric wt % of hydrogen is 9.80.**

3 Sc atoms are placed above the center of the octagon of HGY, and the remaining 3 Sc atoms are placed on the backside of the octagon of HGY. Although the binding energy of the Sc atom is the same on the top of the hexagon and octagon positions (-4.56 eV), we have placed the Sc atom only on the top of the octagons for weight percentage calculations to avoid any possibilities of the metal-metal clustering. Since one unit cell of HGY can adsorb 6 Sc atoms and each Sc atom can adsorb 5 $H_2$ molecules, we have found that the gravimetric wt % of hydrogen for Sc decorated HGY system is 9.80 %, which is much higher than the DoE-US requirements (6.5 %).

### 3.5 Practical feasibility of the system

**Calculation of the desorption temperature for reversible use of hydrogen**

The reversible use of the adsorbed hydrogen molecules depends on the desorption temperature. The desorption temperature should be more than the room temperature otherwise; the adsorbed hydrogens will not remain intact during the small thermal fluctuations. We have calculated the average desorption temperature of the hydrogen molecules using the Van't Hoff equation[75]:

$$\boldsymbol{T_d} = \left(\frac{E_b}{k_B}\right)\left(\frac{\Delta S}{R} - \ln P\right)^{-1} \quad (3)$$

Here, $T_d$ and $E_b$ are the average desorption temperature and average binding energy of the adsorbed hydrogen molecules. $k_B$, $\Delta S$, R, and, P are Boltzmann constant, change in entropy of the hydrogen in transition from gas to liquid phase[76], universal gas constant, and atmospheric pressure, respectively. The calculated value of the desorption temperature is 464 K which is significantly higher than the room temperature and suitable for practical fuel cell applications[77]. Since Sc decorated HGY has suitable adsorption energy, gravimetric weight percentage, and desorption temperature, this system is suitable for fuel cell applications. We have compared some of the crucial parameters of our hydrogen storage system to previously reported systems in **Table 2.**

**Table 2. Hydrogen storage parameters comparison for various carbon nanostructures.**

| Metal decorated systems | Total no. of adsorbed Hydrogen molecules | Average adsorption energy per $H_2$ (eV) | Average desorption temperature (K) | Gravimetric wt % of $H_2$ (%) |
|---|---|---|---|---|
| $B_{40}$ + Y[34] | 5 | -0.211 | 281 | 5.8 |
| $C_{60}$ + Sc[78] | 4 | -0.30 | - | 7.5 |
| SWCNT + Ti[36] | 4 | -0.18 | 230 | 8 |
| SWCNT + Y[40] | 6 | -0.41 | 524 | 6.1 |
| Graphene + Ti[43] | 8 | -0.415 | 511.5 | 7.8 |
| Graphyne + Sc[50] | 4 | -0.60 | | 9.8 |
| Holey graphyne + Li[62] | 4 | -0.22 | 282 | 12.8 |
| **Holey graphyne + Sc (Our work)** | **5** | **-0.36** | **464** | **9.8** |
| **Experimental** | | | | |
| MWCNTs + Pd[55] | - | - | - | 6.0 |
| Graphene + Ni + Al[79] | - | - | - | 5.7 |
| Graphene-Ni Nanocomposites[56] | - | - | - | >6.5 |

The hydrogen uptake for our system is higher than most of the previous reports and the desorption temperature is also more than room temperature and quite suitable for practical fuel cell applications.

**Diffusion energy barrier calculations**

The formation of the metal-metal cluster is one of the major serious concerns in a hydrogen storage system which can decrease the hydrogen uptake to a great extent. Since the binding energy of Sc on HGY is more than the experimental cohesive energy of Sc, the possibilities of metal-metal clustering are negligible in Sc decorated HGY system. To avoid clustering issue, we have placed the Sc atom only above the octagon site keeping the hexagon site vacant. We have also calculated the diffusion energy barrier for the Sc atom in two pathways, as shown in **Fig. 12.**

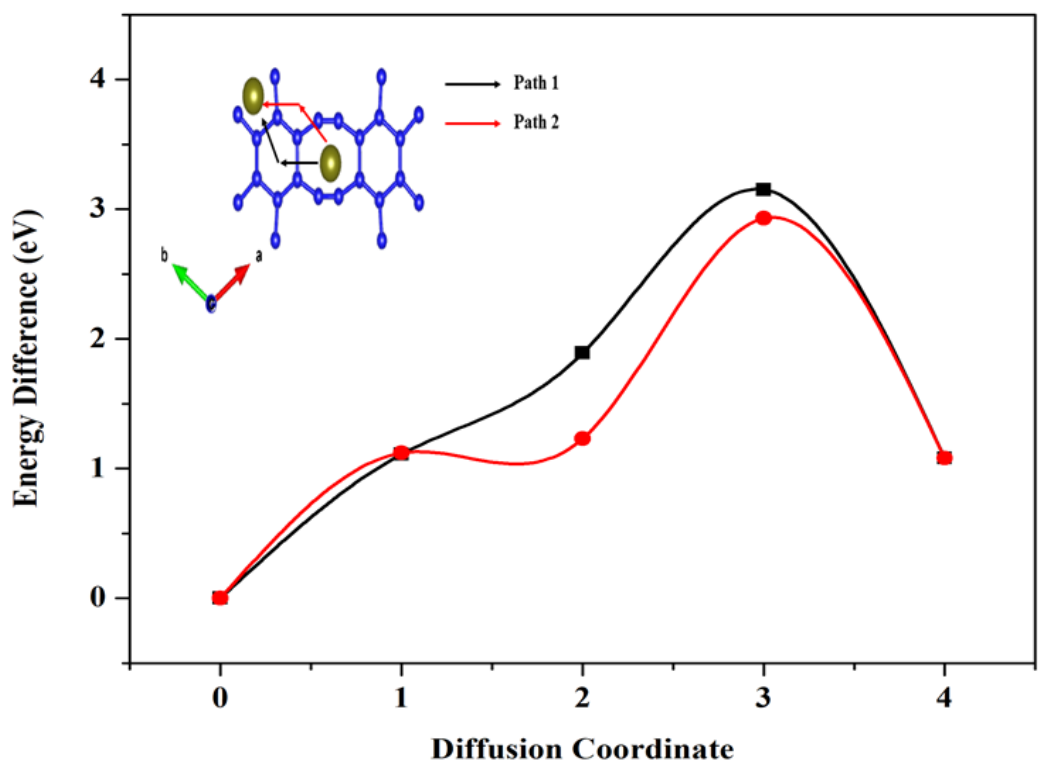

**Fig. 12 Diffusion energy barrier plot for the movement of the Sc-atom along two different pathways. Single point energy difference of current step energy and initial energy is plotted with respect to the small displacements of Sc atom.**

If the diffusion energy barrier of the metal atom is comparable to the thermal energy of the metal atom even at the desorption temperature, then metal clustering may take place. We have calculated the diffusion energy barrier for the movement of the scandium atom from the top of one octagon center to the nearest octagon center. The energy difference of the current displaced configuration and initial configuration is calculated between two stable, relaxed HGY + Sc structures. The diffusion energy barriers for the Sc atom are 3.15 eV and 2.93 eV for path 1 and path 2, respectively. Gao et al.[62] have reported 0.44 eV and 0.26 eV diffusion energy barriers for the Li atom decorated on holey graphyne structure for similar pathways. The presence of a high diffusion energy barrier can restrict the movement of the metal atoms and prevent the system from clustering[62].

Since the average desorption temperature of the absorbed hydrogen molecules is 464 K, we have calculated the thermal energy for the Sc atom at 500 K using the following equation:

$$E = \frac{3}{2} k_B T \quad (4)$$

Here E is the thermal energy, $k_B$ is the Boltzmann constant, and T = 500 K (near desorption temperature). The calculated value of the thermal energy of the Sc atom is 0.065 eV, which is much smaller than the diffusion energy barrier for the Sc atom for path 1 and path 2.

**The solidity of the structures at desorption temperature**

We have investigated the structural integrity of the metal decorated HGY structures at the desorption temperature by performing *ab-initio* molecular dynamics simulations. The metal atom should remain intact with the HGY at the desorption temperature of hydrogen for the reversible hydrogen storage system. Therefore, the molecular dynamics simulations were performed for the HGY + Sc and HGY + 2 Sc system at 500 K. Initially, the metal decorated HGY structures were kept in the microcanonical ensemble for 5 ps time duration and the temperature is increased up to 500 K in the time steps of 1 fs. Next, we have kept the metal decorated HGY structures in the canonical ensemble for another 5 ps at a constant temperature 500 K. The molecular dynamics snapshots of HGY + Sc and HGY + 2 Sc composition at 500 K after 5 ps time duration are shown in **Fig. 13 (a & b).**

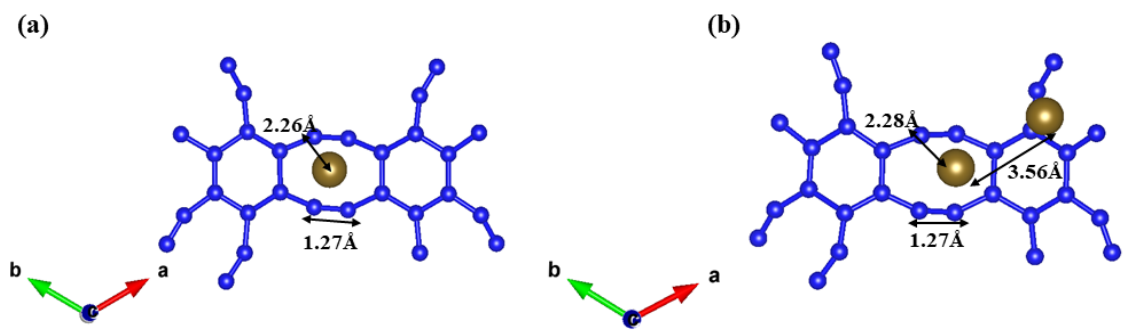

**Fig. 13 (a)** *ab-initio* **Molecular dynamics snapshot of (a) HGY + Sc (b) HGY + 2 Sc after putting the system in canonical ensemble at 500 K for 5 ps.**

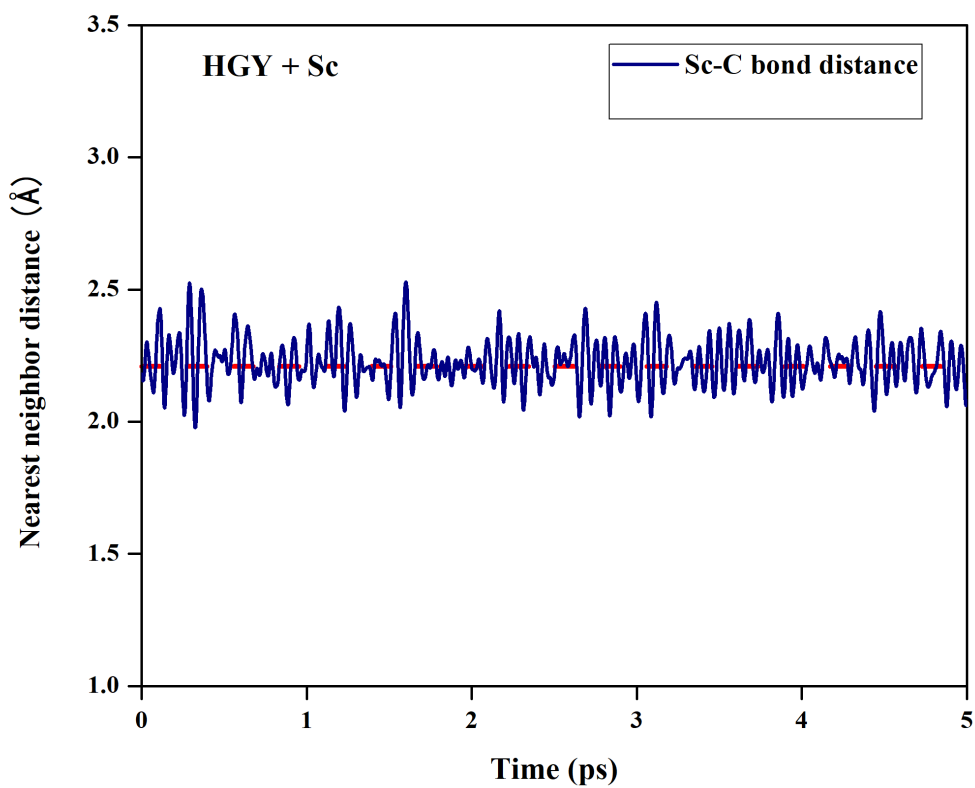

**Fig. 14** Time evolution of the nearest carbon atom of HGY distance from scandium atom at 500 K.

We have also plotted the bond distance between the scandium atom and the nearest carbon of HGY in HGY + Sc system with the time duration of the molecular dynamics simulations, presented in **Fig. 14.** We have observed that the fluctuations in the Sc-C bond distance of HGY are small and oscillating around the mean value Sc-C distance of 2.21 Å. The metal (Sc) remains intact with the HGY at 500 K, and the changes in C-C and C-Sc and Sc-Sc, bond lengths are negligible, implying that our system is practically suitable for hydrogen storage applications.

## 4 CONCLUSIONS

We have presented the hydrogen adsorption and desorption properties of the Sc decorated HGY by using the density functional theory and molecular dynamics simulation. Sc atom is attached on the top of the octagon of the HGY with binding energy -4.56 eV. We have found that each adsorbed Sc atom can bind 5 hydrogen molecules leading to the gravimetric wt % 9.80 for the system, which is much higher than the requirements of the DoE-US. For the reversible use of hydrogen, we have calculated the desorption temperature of the attached hydrogen molecules. The average desorption temperature of the Sc decorated HGY system is 464 K, which is very suitable for practical fuel cell applications. Next, to check the feasibility of the system for practical applications, we have calculated the diffusion energy barrier for the Sc atoms and found that the diffusion energy barriers (3.15 eV and 2.93 eV) are much higher than the thermal energy of the Sc at desorption temperature (0.065 eV). Therefore, we can ensure that the metal-metal clustering will not take place. To check the structural integrity of the system at desorption temperature, we have performed molecular dynamics simulations. We have found that the Sc atoms remain intact to HGY unit cell even at 500 K and the changes in the Sc-C and Sc-Sc bond lengths are small. Therefore, we believe that the Sc decorated HGY is a high-capacity, reversible hydrogen storage device, and our results will motivate the experimentalists to investigate the hydrogen storage capabilities of this system.

**ASSOCIATED CONTENT**

**Supporting information**

# Acknowledgment

VM would like to acknowledge DST-INSPIRE for providing the senior research scholar fellowship and SpaceTime-2 supercomputing facility at IIT Bombay for the computing time. VM would also like to acknowledge Mukesh Singh and Juhee Dewangan for their support. BC would like to thank Dr. T. Shakuntala for support and encouragement. BC also acknowledge support from Dr. S.M. Yusuf and Dr. A. K Mohanty.

# Supporting Information

# High-capacity reversible hydrogen storage in scandium decorated holey graphyne: Theoretical perspectives

*Vikram Mahamiya[a,*], Alok Shukla[a,*], Nandini Garg[b,c], Brahmananda Chakraborty[b,c,*]*

[a]Indian Institute of Technology Bombay, Mumbai 400076, India

[b]High pressure and Synchrotron Radiation Physics Division, Bhabha Atomic Research Centre, Bombay, Mumbai, India-40085

[c]Homi Bhabha National Institute, Mumbai, India-400094

email: vikram.physics@iitb.ac.in ; shukla@phy.iitb.ac.in ; brahma@barc.gov.in

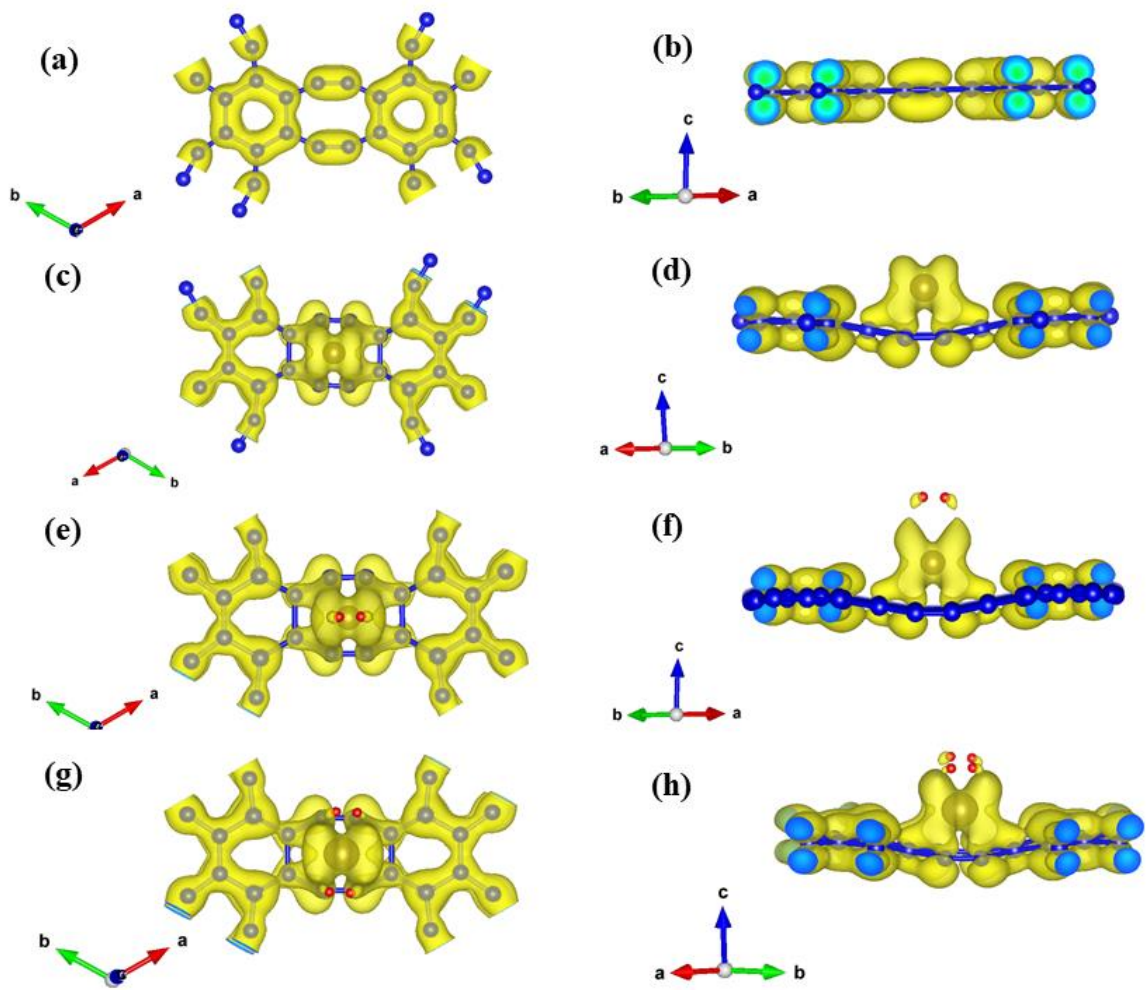

**Fig. S1 (a & b)** Top and side views of the molecular orbital charge density plots of HGY. **(c & d)** Top and side views of the molecular orbital charge density plots of Sc decorated HGY. **(e & f)** Top and side views of the molecular orbital charge density plots of HGY + Sc + 1H$_2$. **(g & h)** Top and side views of the molecular orbital charge density plots of HGY + Sc + 2H$_2$. Here, the blue, golden, and red color corresponds to the C atoms of HGY, Sc atom on the top of the HGY sheet, and H atoms, respectively.

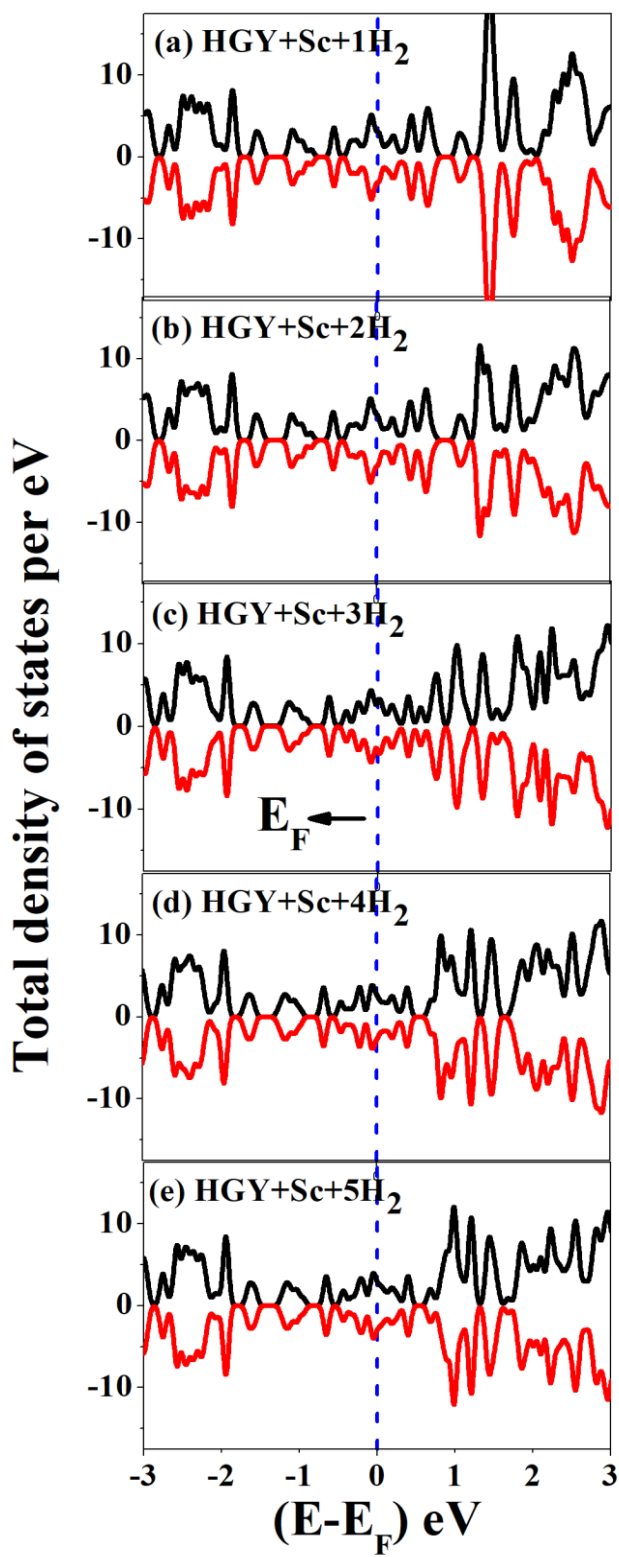

**Fig. S2** Total density of states of (a) HGY + Sc + $1H_2$. (b) HGY + Sc + $2H_2$. (c) HGY + Sc + $3H_2$. (d) HGY + Sc + $4H_2$. (e) (a) HGY + Sc + $5H_2$. Fermi energy is set at zero energy value.